\documentclass{emulateapj}
\usepackage{apjfonts}

\usepackage{hyperref}

\shorttitle{Early NUV absorption at WASP-12b}
\shortauthors{Nichols et al.}

\slugcomment{Submitted 2014 November 25; accepted 2015 February 25}

\begin{document}

\title{Hubble Space Telescope observations of the NUV transit of WASP-12b}

\author{J.~D.~Nichols, G.~A.~Wynn, M.~Goad, R.~D.~Alexander, S.~L.~Casewell, S.~W.~H~Cowley, M.~R.~Burleigh}
\affil{Department of Physics and Astronomy, University of Leicester, Leicester, LE1~7RH, UK}
\email{jdn@ion.le.ac.uk}


\author{ J.~T.~Clarke}
\affil{Center for Space Physics, Boston University, Boston, MA 02215, USA}

\and

\author{D. Bisikalo}
\affil{Institute of Astronomy RAS, 48 Pyatnitskaya Str.,119017, Moscow, Russia}

\begin{abstract}
We present new observations of four closely-spaced NUV transits of the hot Jupiter-like exoplanet WASP-12b using HST/COS, significantly increasing the phase resolution of the observed NUV light curve relative to previous observations, while minimising the temporal variation of the system.  We observe significant excess NUV absorption during the transit, with mean normalised in-transit fluxes of $F_\mathrm{norm}\simeq0.97$, i.e.\ $\simeq$2-5 $\sigma$ deeper than the optical transit level of $\simeq0.986$ for a uniform stellar disk (the exact confidence level depending on the normalisation method used). We further observe an asymmetric transit shape, such that the post-conjunction fluxes are overall $\simeq$2-3 $\sigma$ higher than pre-conjunction values, and characterised by rapid variations in count rate between the pre-conjunction and out of transit levels.  We do not find evidence for an early ingress to the NUV transit as suggested by earlier HST observations.  However, we show that the NUV count rate observed prior to the optical transit is highly variable, but overall $\simeq$2.2-3.0 $\sigma$ below the post-transit values and comparable in depth to the optical transit, possibly forming a variable region of NUV absorption from at least phase $\phi\simeq$0.83, limited by the data coverage.

\end{abstract}


\section{Introduction}

WASP-12b is a transiting hot Jupiter-like exoplanet orbiting a late-F/early-G main sequence star with an orbital semi-major axis of $\simeq$0.0229~AU and an orbital period of $\simeq$1.09~d \citep{hebb09a}.  The radius of the planet $R_p\simeq1.74~\mathrm{R_J}$ is comparable to the planet's Roche lobe radius of  $\simeq1.37R_p$, which has led to the suggestion that the planet may be losing mass through evaporating material overflowing the Roche lobe, as for HD209458b \citep[e.g.][]{VidalMadjar:2003bl,VidalMadjar:2008dz}.  With this possible evaporation in mind, \emph{Hubble Space Telescope} (HST) observations of the near-UV (NUV) transit were obtained by \cite{fossati10a} and \cite{haswell:2012aa} (hereafter F10 and H12, respectively, and F-H together), who observed two transits of the planet in Oct 2009 and Mar 2010, with the aim of detecting elements within the planet's atmosphere.  F10 presented the data from the first of these observing intervals (visits), which comprised 5 contiguous orbits of Cosmic Origins Spectrograph (COS) observations covering the planet's primary transit, as well as modestly before and after.  The NUV observations (2539-2811~\AA) contain a strong Mg \textsc{ii} resonance doublet at $\simeq$2800~\AA, but the stellar continuum was observed to be significantly absorbed across all three COS stripes owing to a blend of thousands of lines of metals including Mg, Na, Fe, Al, Co, Al and Mn, such that no unabsorbed continuum was observed. A key result, initially presented by F10 and confirmed by H12, was an apparent early NUV ingress when compared to optical transit observations. The NUV flux was observed to be lower than the expected out of transit (OOT) level by $\simeq$2$\sigma$ at phase $\phi\simeq0.92$ (note that the optical ingress begins at $\phi\simeq0.94$, see Fig.~2 in F10).  To date there are two main hypotheses put forward to explain the claimed early ingress in the transit of WASP-12b: (a) material of planetary origin overflowing the Roche lobe at the L-1 point and deflecting in the direction of planetary motion \citep{lai10a, bisikalo13a, bisikalo13b, bisikalo14a, cherenkov14a}, and (b) shocked material in a magnetosheath upstream (with respect to the planet's orbit) of the planet's magnetosphere \citep{lai10a, vidotto10a, vidotto11b, vidotto11c, llama11a, alexander15a},  although the data were unable to distinguish between these two scenarios.  The second visit, presented by H12, also exhibited low fluxes at pre-transit phases out to $\phi\simeq0.83$, although there were significant differences between the values obtained in the first and second visits at similar phases.  Considering the $\simeq$6-month separation of the visits and the inherent variability of conditions in stellar and planetary space environments, this is perhaps unsurprising and renders a robust interpretation of the superposed data somewhat difficult.  We have thus observed 4 closely-spaced NUV transits of WASP-12b using HST/COS, in order to build up a light curve with higher phase resolution, while minimising the temporal variation of the system, in an effort to examine more closely the transit NUV light curve.

\section{Data} 
\label{sec:data}

\begin{figure}
\noindent\includegraphics[width=90mm]{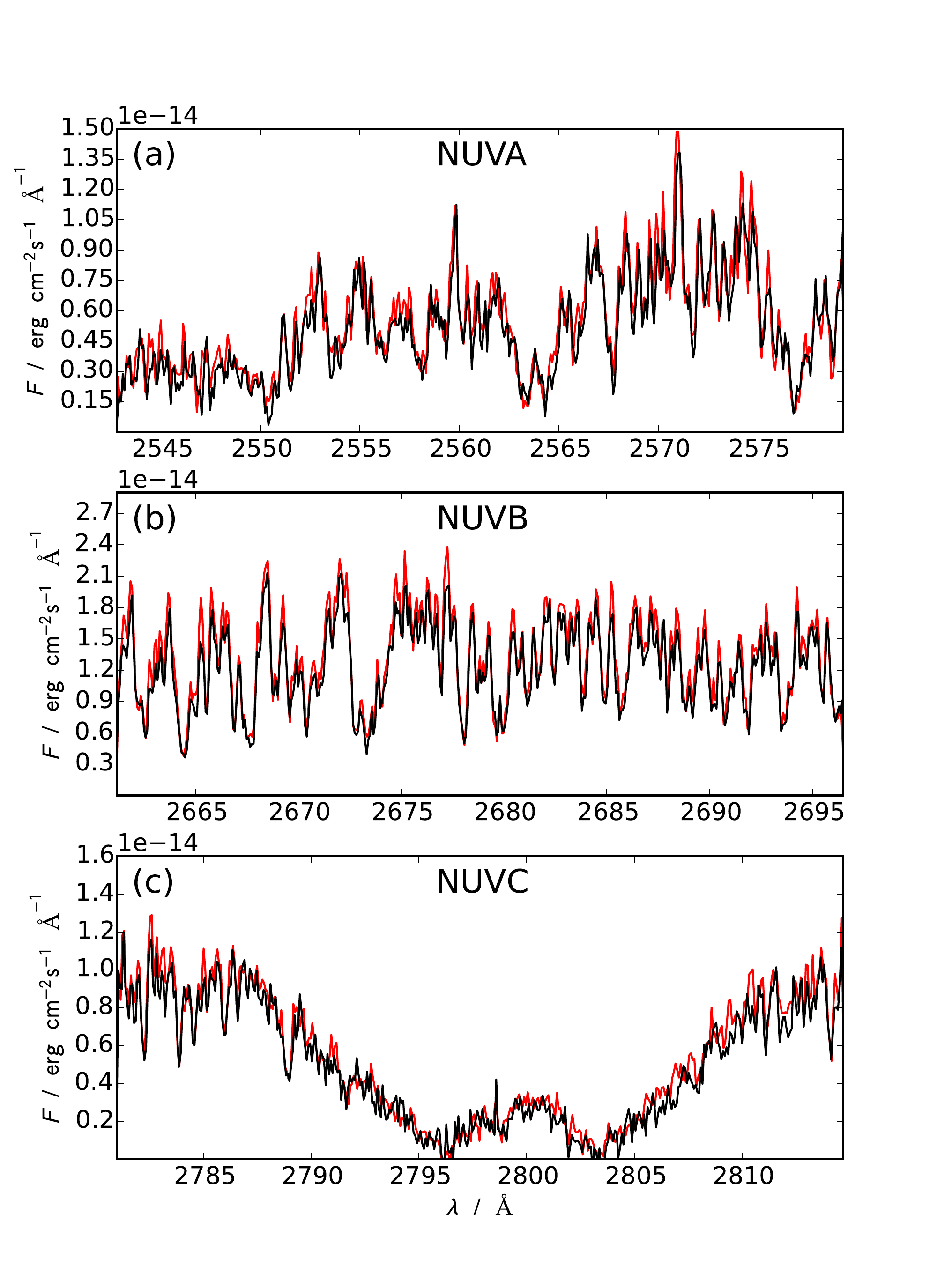}
\caption{  
Figure showing the grand sums of the spectra observed in the three NUV stripes over the program, versus wavelength $\lambda$ in \AA, over the common wavelength envelope for the visits, i.e.\ 2543-2579~\AA\ for NUVA, 2661-2697~\AA\ for NUVB, and 2781-2815~\AA\ for NUVC.  The black line shows the data from this program, while the red shows that obtained by F-H.  The spectra have been averaged over two element bins to increase signal to noise for presentation.
}
\label{fig:specs}
\end{figure}

We observed WASP-12b in late 2013 using HST/COS, a slitless spectrograph with a 2.5 arcsecond aperture \citep{green12a}.  We observed three adjacent transits on 2013 November 27, 28, and 29 in visits W6, W5, and W7 plus a further one, W3, a few weeks earlier on  October 23 (this split timing and numbering results from a guide star failure during the initial execution). The observations were obtained at interleaved phases in order to maximise the phase coverage.  We employed the G285M grating at the 2676~\AA{} setting, which admits NUV light in three non-contiguous wavebands (`stripes' NUVA, NUB, and NUVC) over $\simeq$2500-2800~\AA, with $R\simeq20,000$. The data were reduced using \textsc{CALCOS} v2.19, configured to centre the extraction box over the peak in the cross-dispersion direction, usually only a few pixels away from the nominal position, with the background stripes shifted by the same displacement from the nominal location.  We employed 4 offset wavelength ranges during the program to reduce fixed-pattern noise on the detector (`FP-POS' settings), although we note that we kept the FP-POS settings fixed during each visit, i.e.\ across each transit.  The mean count rates detected are $\simeq$5, $\simeq$14, and $\simeq$6~$\mathrm{counts\;s^{-1}}$ for the NUVA, NUVB, and NUVC stripes respectively, $\simeq$50\% of the rate observed by F10 owing to the $\simeq$12\% / yr decline in sensitivity of COS/NUV/G285M spectroscopy discussed by \cite{osten11a}. The spectra summed over all visits (within the common wavelength envelope for the different FP-POS settings) are shown by the black lines in Figure~\ref{fig:specs}.  We also note that we have only used pixels with good data quality flags.  For comparison, the red lines show the spectra obtained by F10, with which our data are consistent.  Note that in both cases the centre-of-mass velocity of 19.06~km~s$^{-1}$ observed for WASP-12 by \cite{husnoo11a} is corrected for, and that in this plot the spectra are $\simeq$1~\AA~longward of those plotted by F-H, who shifted their spectra to fit to a stellar template.  We further note that a significant redward shift was observed between visit W7 and the others, determined by cross-correlation to be $\simeq$0.3~\AA, although it is not obvious whether this is a shift in wavelength owing to a calibration uncertainty, or a real change in velocity (corresponding to $\simeq$30~km~s$^{-1}$).  As observed previously, stripe NUVC is dominated by a strong Mg \textsc{ii} resonance doublet, but it is important to note that, as discussed by F-H, the stellar continuum is significantly absorbed in all three stripes owing to the complex blend of metal lines, such that no unabsorbed continuum is observed at any phase.  In this Letter we concentrate on the light curves obtained by summing the counts obtained over these wavebands. \\

\section{Analysis} 
\label{sec:analysis}

\begin{figure}
\noindent\includegraphics[width=90mm]{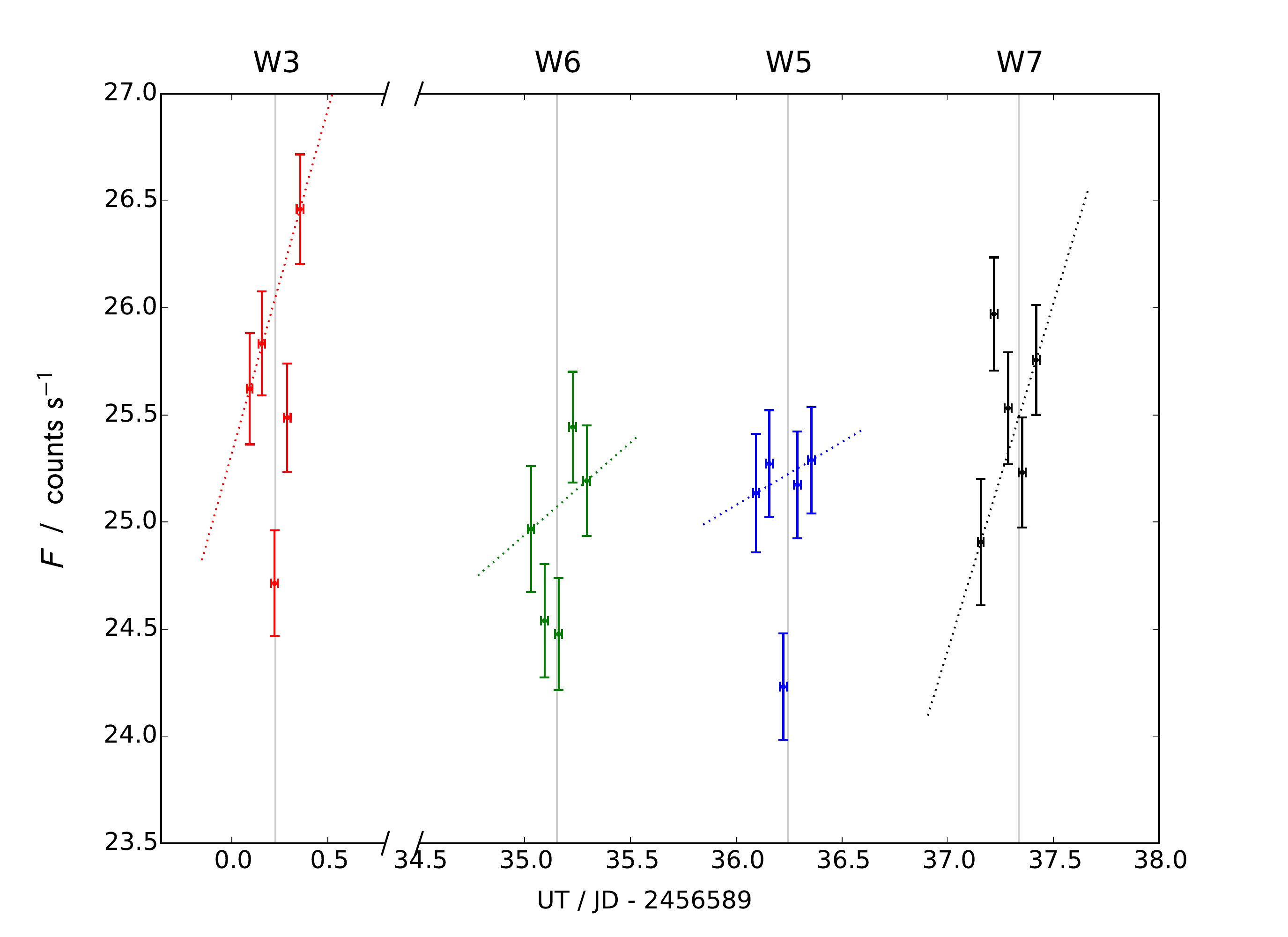}
\caption{
Plot showing the flux $F$ in $\mathrm{counts\;s^{-1}}$ versus UT (JD - 2456589).  The data from visits W3, W6, W5, and W7 are indicated by red, green, blue, and black points, respectively. The error bars are computed as described in the text.  The  dotted lines indicate the slopes of the linear fit used for the linear normalisation, color coded for each visit as described above.  The vertical grey lines indicate the transit mid-points.  
}
\label{fig:fluxtime}
\end{figure}

\begin{figure}[b]
\noindent\includegraphics[width=90mm]{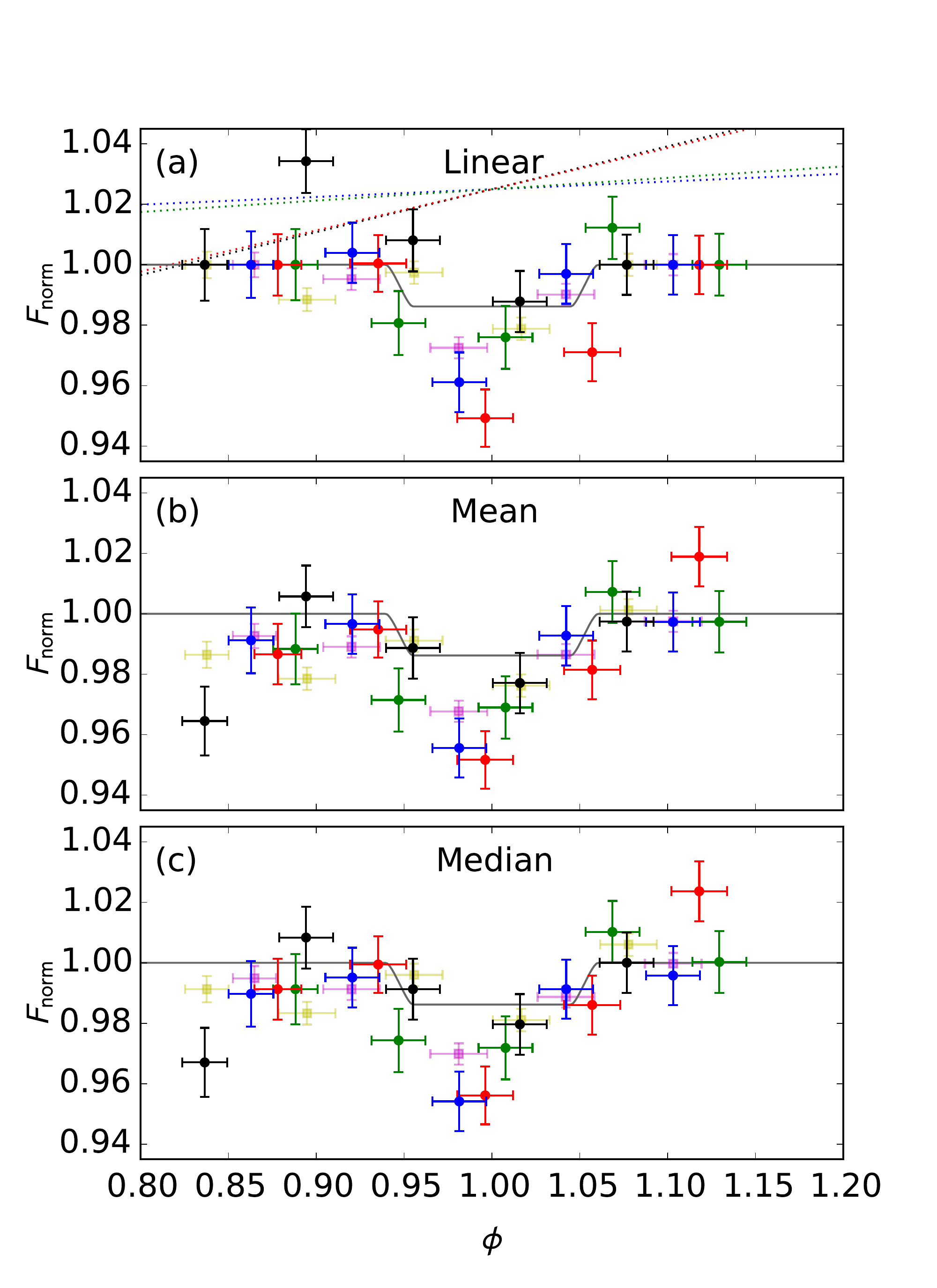}
\caption{
Plot showing the normalised flux $F_\mathrm{norm}$ versus phase $\phi$ for (a) the linear, (b) mean, and (c) median methods of normalisation as discussed in the text.  The data from visits W7, W5, W3, and W6 are indicated by black, blue, red, and green points, respectively, while visits 1 and 2 of F-H are shown by the faded yellow and magenta square points. The error bars are computed as described in the text.  The grey solid line indicates the optical transit for a disk of uniform brightness given by Equation~1 of \cite{mandel02a}.  The  dotted lines in panel (a) indicate the slopes of the linear fit used for normalisation, color coded for each visit as described above.
}
\label{fig:lcsums}
\end{figure}

We first show in Fig.~\ref{fig:fluxtime} the total observed NUV fluxes $F$ in counts s$^{-1}$ for each HST orbit versus universal time (UT).   The vertical error bars are the sum in quadrature of the errors calculated in the pipeline reduction, dominated in practice by Poisson statistics, and the horizontal error bars indicate the duration of each exposure.  The data from visits W3, W6, W5, and W7 are indicated by red, green, blue, and black points, respectively.  It is apparent that there is variation in the overall count rates observed between each visit, and in general observed fluxes $F$ in counts s$^{-1}$ are normalised to some value $F_0$, such that normalised flux $F_\mathrm{norm}=(F/F_0)$.  It is standard practice to normalise to a linear fit to the out-of-transit (OOT) points before and after the transit, in order to de-trend any slowly-varying changes in stellar activity during the observation interval.  This technique was employed by F-H, and we thus show using dotted lines in Fig.~\ref{fig:fluxtime} such fits to the first and last points of each visit, and in Figure~\ref{fig:lcsums}a we plot the phase-folded normalised flux $F_\mathrm{norm}$, versus planet orbital phase $\phi$.  Our data are color-coded as for Fig.~\ref{fig:fluxtime}, while the data obtained by F-H in their visits 1 and 2 are also shown by the faded yellow and magenta square points.  We show, using dotted lines, the gradients of the linear fits as shown in Fig.~\ref{fig:fluxtime}, colour coded as above. The error bars are computed  as for Fig.~\ref{fig:fluxtime}. We note that the errors include a substantial contribution from background counts, which increased by a factor of $\simeq$5 between 2009 and 2013 \citep{roman-duval14a} and which, combined with the loss in sensitivity, yield uncertainties larger than those for the F-H data by a factor of $\simeq$3.  We also note that the error bars in F-H are consistent with the omission of background counts in their error analysis, implying that they are underestimated by factors of $\simeq$1.3-1.8.  The grey solid line indicates the optical transit for a disk of uniform brightness given by Equation~1 of \cite{mandel02a} for comparison.  The transit ephemerides employed are those used by H12, which we verified are still valid by observation of the optical transit on 2013 November 30, i.e.\ subsequent to those observed using HST, using the ground-based 2m telescope with an \emph{R}-band filter at Terskol, Russia.  Variability and scatter notwithstanding, the transit is observed with a mean normalised flux of $F_\mathrm{norm}=0.969\pm0.007$ for the four (blue, red, green, and black) near-mid-transit points, with the uncertainty calculated by propagating the individual error estimates. The values of the latter are essentially identical to those obtained by taking the standard error of the mean indicating that the scatter is indeed dominated by Poisson statistics. The observed transit is thus $\simeq$2.4$\sigma$ lower than the value of $\simeq$0.986 for the depth of the optical transit, i.e.\  consistent with the observations of F-H. Having said that, there is no evidence in Fig.~\ref{fig:lcsums}a of an early ingress such as that proposed previously by F-H; of the four new data points around ingress, only that of visit W6 is below the model transit level, by $\simeq1.2$$\sigma$.\\

However, the linear normalisation used to produce Fig.~\ref{fig:lcsums}a may not be an appropriate technique to apply to NUV data, for a number of reasons.  First, the NUV provides an opportunity to observe absorption by matter in the space environment surrounding the planet, as discussed by e.g.\ \cite{lai10a, bisikalo13a, vidotto10a,llama11a,alexander15a}.  Hence, deviation from the `true' OOT level prior to the optical transit might be expected, and normalising these points to $F_\mathrm{norm}=1$ would remove observable absorption signatures of the space environment.  Second, as discussed below, the OOT fluxes at phases prior to the optical ingress are consistently below those post-egress, such that the observed gradients are all positive, which is inconsistent with quasi-random stellar variability and inconsistent with the observed variation between visits shown in Fig.~\ref{fig:fluxtime}.  Third, for two visits the gradient of the linear fit is considerable, being affected by outlying fluxes observed in the first or last orbits. In Figures~\ref{fig:lcsums}b-c and Figures thereafter we thus show the results from two other methods of normalisation.  These are normalisation to (Fig.~\ref{fig:lcsums}b) the mean flux computed separately for each visit, and (Fig.~\ref{fig:lcsums}c) the median flux determined separately for each visit.  In each case a further constant multiplicative factor is thereafter applied in order to normalise the mean of all the post-optical transit data to 1.  In the case of the median light curve this mean value omits an anomalously high outlier point, since the median normalisation procedure is intended to remove the influence of these high points, which act to raise the mean and shift the remaining points down.  It is worth noting that the scatter in the median light curve discussed below is indeed modestly reduced over that which uses the mean.\\

As for Fig.~\ref{fig:lcsums}a, an over-deep transit is observed in Figs.~\ref{fig:lcsums}b and c, with mean fluxes of the in-transit points of $F_\mathrm{norm}=0.966\pm0.004$ and $0.969\pm0.004$, respectively, i.e.\ $\simeq$5.1$\sigma$ and $\simeq$4.1$\sigma$ deeper than the optical transit.  The transit is asymmetric; the mean of the pre-conjunction in-transit fluxes is $F_\mathrm{norm}=0.958\pm0.005$ ($0.960\pm0.005$) for the mean (median) normalisation, while that for the post-conjunction in-transit fluxes is $F_\mathrm{norm}=0.974\pm0.005$ ($0.977\pm0.005$).  Again, we do not observe robust evidence of an early ingress in these data, since none of the points are significantly below the model transit curve, except for visit W6, which deviates by $\simeq2.1$$\sigma$ ($\simeq1.8$$\sigma$) for the mean (median) normalisation methods. It is also worth noting that the pre-transit points are, in general, below the OOT level as defined above and, including the F-H data, there are 7 and 5 pre-transit points whose values are $\leq1\sigma$ below the OOT level for the mean and median normalisation methods, respectively.  These low fluxes do not comprise an early ingress of the form described previously by F10, but rather a broad region of enhanced absorption at least out to $\phi\simeq$0.83.  There is a possible positive gradient in flux with phase in the pre-transit region in our data; the gradients of linear fits to these pre-transit (only) points are $0.19\pm0.13$ ($0.16\pm0.13$) of normalised flux $F_\mathrm{norm}$ per unit phase for the mean (median) normalisations. \\

\begin{figure}
\noindent\includegraphics[width=90mm]{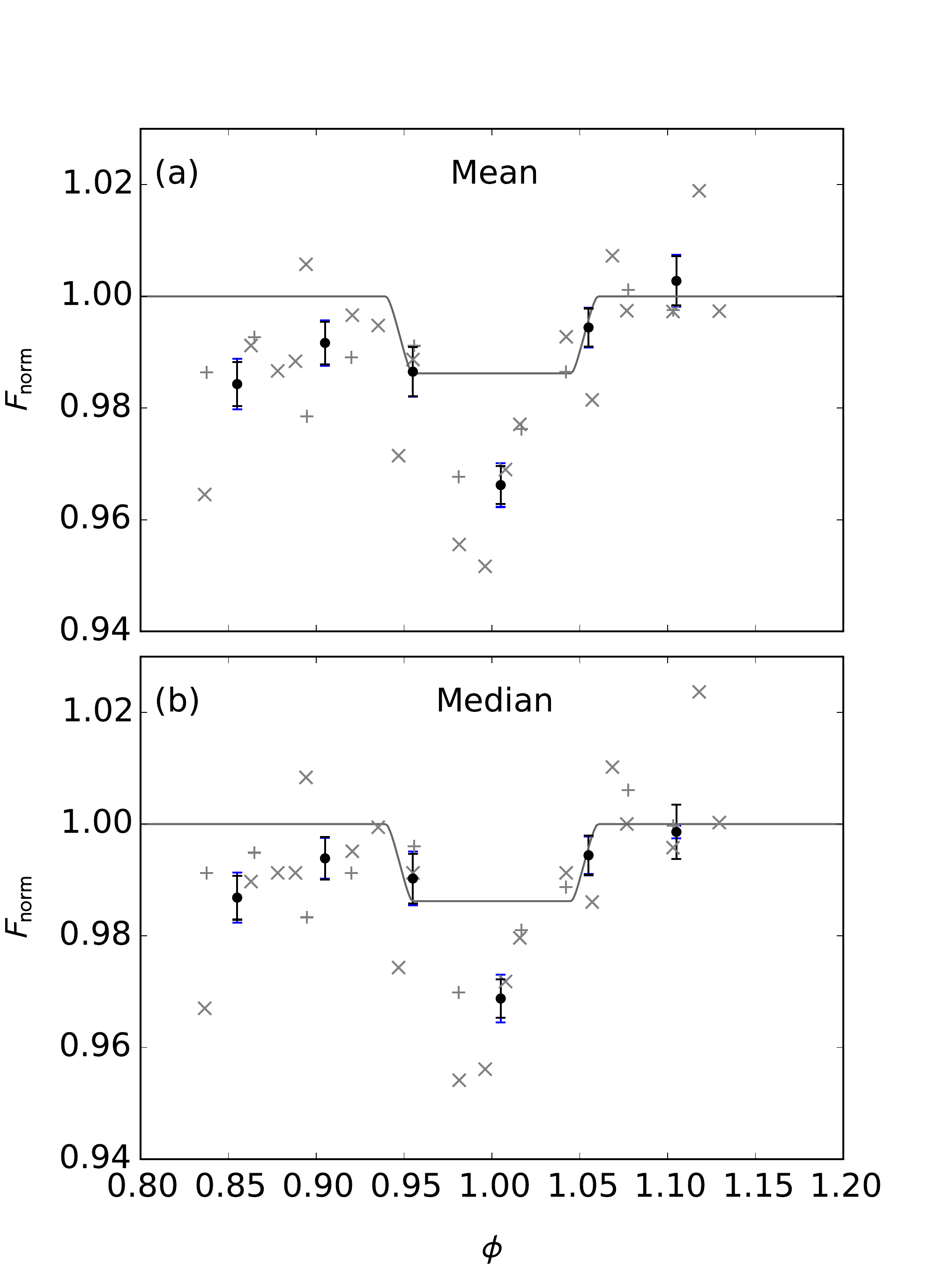}
\caption{
Plot showing the phase binned normalised flux $F_\mathrm{norm}$ versus $\phi$ for (a) the mean, and (b) median methods of normalisation.  The black points show the flux averaged over 0.05 bins of phase, while the grey points show the data from Fig.~\ref{fig:lcsums}b and c for comparison (crosses and pluses indicate our and F-H data points, respectively).   Error bars are computed as discussed in the text, and the optical transit is shown by the grey line. 
}
\label{fig:binned}
\end{figure}

We show in Figure~\ref{fig:binned} the light curve of the combined data sets averaged over 0.05-wide bins of phase.  The individual data points are shown by the grey crosses (our data) and pluses (F-H data), and the binned points are shown by the circles, which also indicate the propagated errors (black error bars) and standard errors of the bin means (blue error bars).  Note that, in keeping with the above discussion, the results for the median curve do not include the significant outlier point at $\phi\simeq$1.12.  The pre-transit binned points are below the OOT level by $\simeq$$1.6-4.0\sigma$. 
Overall, the mean of all the pre-transit points is $F_\mathrm{norm}=0.988\pm0.003$ ($0.990\pm0.003$) for the mean (median) normalisation, where the uncertainties given are whichever is the larger of either the propagated error estimate or the standard error of the mean.  These correspond to deviations of $\simeq$$3.7\sigma$ ($\simeq$$3.1\sigma$) from unity, and the overall differences between the pre- and post-transit points are $\Delta F_\mathrm{norm}=0.014\pm0.005$ ($0.010\pm0.005$), where the uncertainties are given by the sum in quadrature of the errors associated with the respective mean values.  The pre-transit flux is thus overall lower than the post-transit flux by $\simeq$$2.2-3.0\sigma$, and is comparable in depth to the optical transit.  \\

\begin{figure}[b]
\noindent\includegraphics[width=90mm]{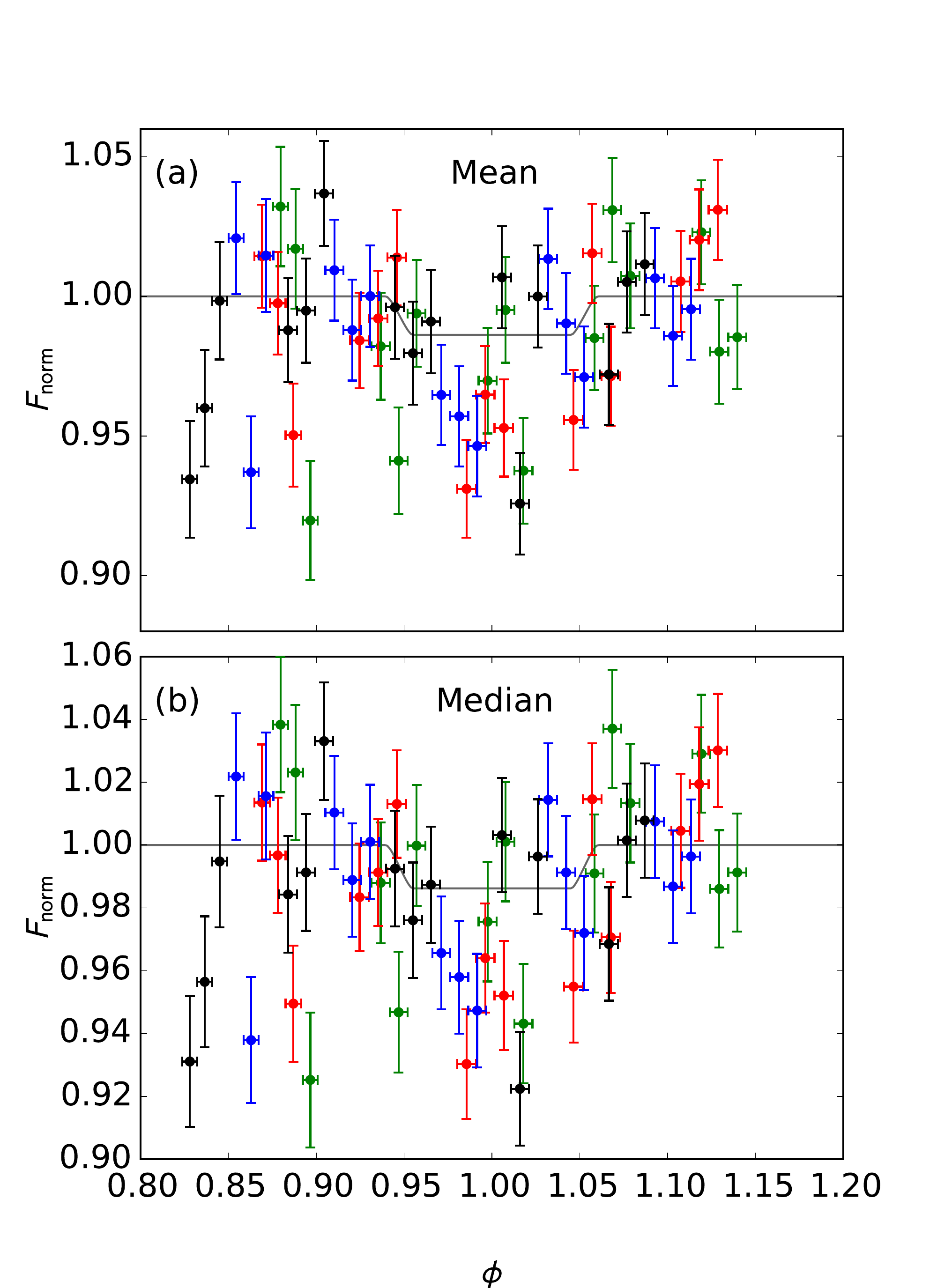}
\caption{
As for Fig.~\ref{fig:lcsums}, but with each data exposure split into three data points for the mean and median normalisation methods. 
}
\label{fig:lcsplit}
\end{figure}

We finally show in Fig.~\ref{fig:lcsplit} the light curve with each time-tagged exposure split into three points, significantly increasing the size of the error bars, but providing illuminating details nonetheless.  We first note that the transit asymmetry discussed above arises due to rapid sub-exposure variations of the count rate over phases $1.01\lesssim \phi\lesssim 1.05$, in which three visits exhibit at least one data point $\simeq$2-3$\sigma$ above the pre-conjunction mean of $F_\mathrm{norm}\simeq0.96$, and similar to the OOT level.  However, the flux is not uniformly high in this region, with two points in visits W6 and W7 exhibiting values similar to the pre-conjunction values, such that the light-curve is somewhat complicated in this region.  Considering also the pre-transit points, we note that again, there are rapid sub-exposure changes in count rate, and each visit exhibits at least one point significantly below the others at the OOT level.

\section{Discussion} 
\label{sec:discussion}

First, these observations confirm the existence of significant, sustained excess NUV absorption during the transit of WASP-12b, although we obtained no convincing evidence for an early ingress of the form discussed by F10.  We have further observed that the NUV transit is asymmetric, with significantly higher fluxes observed during the post-conjunction transit than during the pre-conjunction phases.  This arises owing to sub-exposure timescale variations between deep values and fluxes comparable to the OOT level, during this interval.  In analysing these data we have considered, and ruled out, a number of possible sources of systematic error, such as count rate-varying photometric sensitivity of COS, pointing errors, FP-POS settings, exposure times, and background removal effects.  While the photometric stability of the first orbit of each visit may be reduced by systematic effects (discussed below), it seems less likely that the 3rd or 4th orbit of each visit would be impacted in this way.  Similarly, it is possible, but perhaps unlikely, that stellar flares would brighten at the same planetary orbital phase. \\

The early low fluxes occur predominately during the first orbit of each visit, which for some instruments onboard HST (e.g.\ STIS) have proved photometrically unstable.  There is, however, no obvious trend in Fig.~\ref{fig:lcsplit} for e.g.\ increasing flux with phase during these exposures, and the first orbit of all but one visit (W7) has highest flux in the first split data point.  In each exposure there is usually one data point significantly below the other two.  It is thus not obvious whether these early phase rapid variations in flux are due to NUV absorption that is variable on sub-exposure timescales, or unstable photometry, such that the observation of low pre-transit fluxes should be considered as tentative.  It is also worth noting that the continuing deterioration of the COS/NUV detector is such that similar future observations of WASP-12b are not likely to yield further insight on this matter.\\

If the reduced fluxes are not instrumental, as discussed by H12 absorption out to $\phi\gtrsim0.83$ would require any absorbing structure associated with the planet to extend to at least $\simeq$30~R$\mathrm{_p}$ ahead of the planet in its orbit, i.e.\ considerably larger than the lower limit of $\simeq$4~R$\mathrm{_p}$ determined by \cite{lai10a} using the F10 data.  In the models of \cite{bisikalo13a, bisikalo13b, bisikalo14a} and \cite{cherenkov14a}, planets undergoing Roche lobe overflow exhibit either `quasi-closed' or `open' atmospheres, dependent on the rate of mass outflow from the planet.  For WASP-12b the prominence of a quasi-closed atmosphere could extend to $\simeq$15-30~R$\mathrm{_p}$ depending on the shape of the outflowing stream, although this would require extremely low stellar wind speeds (of order $\simeq$10 $\mathrm{km\;s^{-1}}$).  The limiting planetary mass outflow rate for an open atmosphere was estimated by \cite{cherenkov14a} (albeit for HD 209458b) to be $\simeq$$3\times 10^{10}\;\mathrm{g\,s^{-1}}$, i.e.\  an order of magnitude greater than the limit for a quasi-closed atmosphere configuration, such as that considered for WASP-12b by \cite{bisikalo13a}.  Further modelling of the gas dynamic interaction at WASP-12b will determine the parameters required to create such a large obstacle via Roche lobe overflow at this planet.  In the case of absorption by shocked stellar wind material in a magnetosheath, a magnetosphere with a standoff radius of 6-9~R$\mathrm{_p}$ would produce a broad, low-Mach shock that extends to this distance as projected ahead of the planet, and which would provide the required NUV absorption if the optical depth of the sheath material is $\tau\simeq0.1$ \citep{alexander15a}. A magnetosphere of this radius may be produced by a dipolar planetary field of surface equatorial strength $\simeq$2-25~G, where the range reflects the various assumptions that can be made about the magnetic, plasma and dynamic pressures either side of the magnetopause boundary, as discussed further by \cite{alexander15a}.  It is finally worth noting that the light curves modelled by \cite{alexander15a} are asymmetric with a form qualitatively similar to that observed, although observations with higher signal-to-noise of a closer hot Jupiter, e.g.\ WASP-18b, are required to constrain this aspect of the NUV transit.\\

\section{Summary} 
\label{sec:summary}

We have presented new COS/HST NUV observations in and around four closely-spaced transits of exoplanet WASP-12b. We have confirmed the existence of reproducible excess NUV absorption during the planet's transit at a level $\simeq$2-5 $\sigma$ deeper than the optical transit level, and we showed that the transit is apparently asymmetric, with the post-conjunction in-transit fluxes significantly larger overall than the pre-conjunction values, and characterised by rapid variations.  We do not find strong evidence for an early ingress previously suggested by F10, however we have also presented tentative evidence of a region of variable NUV absorption from at least $\phi\simeq$0.83 (limited by the data coverage), with orbit-integrated pre-transit fluxes comparable in depth to the optical transit.  While systematic effects cannot be ruled out, such absorption could be produced physically by a low Mach stellar wind interaction with a planetary magnetosphere, or rapid flow over the Roche lobe.  Further observations of the NUV transits of hot-Jupiters over wider phase ranges than those obtained to date are required to constrain form of the pre-transit light curve and distinguish between these scenarios.  \\

\acknowledgments{This work is based on observations made with the NASA/ESA Hubble Space Telescope, obtained at the Space Telescope Science Institute, which is operated by AURA, Inc.\ for NASA. JDN was supported by an STFC Advanced Fellowship (ST/I004084/1). RDA acknowledges support from STFC through an Advanced Fellowship (ST/G00711X/1), and from the Leverhulme Trust through a Philip Leverhulme Prize. GW, MG, SWHC, RDA, and MRB were supported  by STFC Grant ST/K001000/1. DB was supported by the RSCF (Project 14-12-01048). JDN thanks J.~Ely for useful discussions regarding COS systematics and the wavelength calibration.}

\clearpage

\end{document}